\documentstyle[12pt]{article}
\input epsf
\def \b{{\cal B}}
\def \beq{\begin{equation}}
\def \eeq{\end{equation}}
\def \beqn{\begin{eqnarray}}
\def \eeqn{\end{eqnarray}}
\def \cn{Collaboration}

\def \s{\sqrt{2}}

\def \v#1#2{V_{#1#2}}

\begin{document}
\renewcommand{\thetable}{\Roman{table}}
\rightline{EFI-99-10}
\rightline{hep-ph/9903543}
\rightline{March 1999}
\bigskip
\bigskip
\centerline{{\bf ON LARGE FINAL-STATE PHASES IN HEAVY MESON DECAYS}
\footnote{To be submitted to Phys.~Rev.~D.}}
\bigskip
\centerline{\it Jonathan L. Rosner}
\centerline{\it Enrico Fermi Institute and Department of Physics}
\centerline{\it University of Chicago, Chicago, IL 60637}
\bigskip
\centerline{\bf ABSTRACT}
\medskip
\begin{quote}

An attempt is made to identify circumstances under which the weak decays of $D$
and $B$ mesons may display large differences between eigenphases of strong
final-state interactions.  There are several cases in which rescattering from
other final states appears to enhance decay rates with respect to estimates
based on the factorization hypothesis. 

\end{quote}
\medskip
\leftline{\qquad PACS codes:  13.25.-k, 11.30.Hv, 12.40.Nn, 13.75.Lb}
% \newpage
\bigskip

\centerline{\bf I. INTRODUCTION}
\bigskip

Ever since the discovery of CP violation in neutral kaon decays \cite{CCFT},
attempts have been made to learn more about its origin. The ratios $R_{+-} =
\Gamma(K_L \to \pi^+ \pi^-)/\Gamma(K_S \to \pi^+ \pi^-)$ and $R_{00} =
\Gamma(K_L \to \pi^0 \pi^0)/\Gamma(K_S \to \pi^0 \pi^0)$ are predicted to be
equal in any model (such as a superweak \cite{SW} one) in which CP violation
arises purely via $K^0$--$\bar K^0$ mixing, but can differ from one another by
up to {\cal O}(1\%) \cite{BJL} in the Kobayashi-Maskawa (KM) theory \cite{KM}
based on phases in weak coupling constants. 

The two most recent previous measurements of $\Delta R \equiv (R_{+-}/R_{00}) -
1$: $(0.44 \pm 0.35)\%$ (Fermilab~E731 \cite{E731}) and $(1.38 \pm 0.39)\%$
(CERN~NA31 \cite{NA31}) have now been joined by those of a new experiment with
more compelling statistics, which finds $\Delta R = (1.68 \pm 0.25)\%$
(Fermilab~E832 \cite{E832}). Superweak models are ruled out.  The effect is
near the upper limit of theoretical estimates \cite{BJL}, but can be
accommodated by reasonable values of hadronic matrix elements and strange quark
mass. The new result will reduce the uncertainty on the parameters of the
Cabibbo-Kobayashi-Maskawa (CKM) matrix \cite{KM,Cab} describing the weak
charge-changing couplings of quarks. 

A key test of the KM theory involves decays of $B$ mesons (containing $b$
quarks). CP violation can manifest itself in as follows in such decays:

(1) Decays of neutral $B$ mesons to CP eigenstates such as $J/\psi K_S$ and
$\pi \pi$ can directly probe CKM phases, since their interpretation is
generally immune to questions of strong final-state interactions.  However,
such studies require identification of the flavor ($B^0 = \bar b d$ or $\bar
B^0 = b \bar d$) of the neutral $B$ meson at time of production.  This
requirement can be quite demanding.  It has been addressed in a recent
experiment by the CDF Collaboration \cite{CDFbeta} at the Fermilab
proton-antiproton collider, which finds a difference between the rates for $B^0
\to J/\psi K_S$ and $\bar B^0 \to J/\psi K_S$ at slightly under the $2 \sigma$
level.  Forthcoming electron-positron and hadron studies should prove much more
incisive. 

(2) Decays of $B$ mesons to ``self-tagging'' final states $f$, in which one can
distinguish $f$ (e.g., $K^+ \pi^-$) from its CP-conjugate $\bar f$ (e.g., $K^-
\pi^+$) can manifest a CP-violating asymmetry if there are two decay channels
characterized by differing weak phases $\phi_{1,2}$ and strong phases
$\delta_{1,2}$.  Writing the decay amplitudes as 
\beq
{\cal A}(B \to f) = a_1 e^{i \phi_1} e^{i \delta_1}
           + a_2 e^{i \phi_2} e^{i \delta_2}~~~,
\eeq
\beq
{\cal A}(\bar B \to \bar f) = a_1 e^{-i \phi_1} e^{i \delta_1}
                     + a_2 e^{-i \phi_2} e^{i \delta_2}~~~,
\eeq
we note that the weak phases $\phi_i$ change sign under CP-conjugation, whereas
the strong phases $\delta_i$ do not.  The decay rate asymmetry $A(f)$ is then
given by
\beq \label{eqn:asy}
A(f) \equiv \frac{\Gamma(B \to f) - \Gamma(\bar B \to \bar f)}
                 {\Gamma(B \to f) + \Gamma(\bar B \to \bar f)} =
\frac{2 a_1 a_2 \sin(\phi_1 - \phi_2)\sin(\delta_1 - \delta_2)}
     {a_1^2 + a_2^2}~~~.
\eeq
A non-zero asymmetry of this type requires both the weak phases and the strong
phases to differ from one another in at least two channels.  Whereas it is
straightforward to estimate weak phase differences in typical theories such as
that of Kobayashi and Maskawa, the anticipation of strong phase differences is
much more problematic \cite{BSS,resc}.

In the present paper we examine several instances of large strong phase
differences, in search of a common thread whereby other such cases can be
identified.  We build upon several studies by Suzuki which have identified
large final-state phases in $J/\psi$ \cite{SuzVP,SuzPP} (Sec.~II) and charmed
meson \cite{SuzDB} (Sec.~III) decays.  We conclude that large final-state
phases are a possibility in any process in which a pair of quarks annihilates
hadronically.  Such cases include not only those studied by Suzuki in $J/\psi$
decays, but penguin amplitudes contributing to $b \to s$ processes (Sec.~IV),
including those involving $\eta'$ production.  The case of $B$ decays to
charmed final states, in which large final-state phases do not appear to be
encountered \cite{SuzDB,HNN}, is treated in Sec.~V. 

Although cases with large final-state phases cannot be identified with
certainty, the measurement of $A(f)$ and knowledge of $a_1$ and $a_2$ in
Eq.~(\ref{eqn:asy}) permit one to place a lower bound on $|\sin (\phi_1 -
\phi_2)|$, which can be quite useful in constraining CKM parameters. It is thus
useful to identify promising cases in which the asymmetry $A(f)$ can be large. 
We summarize these cases, noting open experimental questions, in Sec.~VI.
\bigskip
% \newpage

\centerline{\bf II.  CHARMONIUM DECAYS}
\bigskip

\leftline{\bf A.  $J/\psi$ decays}
\bigskip

Recently Suzuki has noted that the three-gluon and single-photon amplitudes in
decays of the form $J/\psi \to VP$ \cite{SuzVP} and $J/\psi \to PP$
\cite{SuzPP} appear to have relative phases of approximately $\pi/2$.  Here and
subsequently $V$ will denote a light vector meson:  $V = (\rho,~\omega,~
K^*,~\phi)$, while $P$ will denote a light pseudoscalar meson:  $P = (\pi,~K,~
\eta,~\eta')$.  Without retracing Suzuki's whole analysis, we review the
essential points, beginning with the three $VP$ decays $J/\psi \to K^{*+} K^-$,
$J/\psi \to K^{*0} \bar K^0$, and $J/\psi \to \omega \pi^0$.  We shall show
that the amplitudes for these three processes form a triangle with significant
area, from which one can infer non-trivial relative phases of strong and
electromagnetic contributions. 

We consider only three-gluon and one-photon contributions, neglecting others
involving (for example) two gluons and one photon and neglecting contributions
from isospin mixing in the neutral pion.  The strong-decay amplitudes for
$J/\psi \to K^{*+} K^-$ and $J/\psi \to K^{*0} \bar K^0$ are equal, while the
decay $J/\psi \to \omega \pi^0$ is isospin-violating and proceeds only
electromagnetically.  The one-photon amplitudes for production of $K^{*+} K^-$,
$K^{*0} \bar K^0$, and $\omega \pi^0$, are proportional respectively to $\mu_u
+ \mu_s$, $\mu_d + \mu_s$, and $\mu_u - \mu_d$, respectively, where $\mu_i =
Q_i|e|/(2 m_i)$ is the magnetic moment of quark $i$ whose charge and mass are
$Q_i$ and $m_i$. 

The three amplitudes of interest can then be written in terms of two parameters
$A$ and $B$ and a quark mass ratio $r = m_{u,d}/m_s$ as
\beq
{\cal A}(J/\psi \to K^{*+} K^-) = A + B \left( \frac{2}{3} - \frac{r}{3}
\right) ~~~, 
\eeq
\beq
{\cal A}(J/\psi \to K^{*0} \bar K^0) = A + B \left( - \frac{1}{3} - \frac{r}{3}
\right)~~~,
\eeq
\beq
{\cal A}(J/\psi \to \omega \pi^0) = B~~~,
\eeq
so that they satisfy a triangle relation
\beq \label{eqn:VPtri}
{\cal A}(J/\psi \to K^{*+} K^-) - {\cal A}(J/\psi \to K^{*0} \bar K^0)
= {\cal A}(J/\psi \to \omega \pi^0)~~~.
\eeq

To estimate the relative amplitudes we use the observed branching ratios
\cite{PDG}: 
$$
{\cal B}(J/\psi \to K^{*+} K^-) = (0.25 \pm 0.02)\%~~,~~~
{\cal B}(J/\psi \to K^{*0} \bar K^0) = (0.21 \pm 0.02)\%~~,~~~
$$
\beq
{\cal B}(J/\psi \to \omega \pi^0) = (0.042 \pm 0.006)\%~~~.
\eeq
We define magnitudes of amplitudes to be the square roots of these branching
ratios.  Kinematic SU(3)-breaking may be included by correcting the $\omega
\pi^0$ amplitude for the slightly larger center-of-mass 3-momentum in $J/\psi
\to \omega \pi^0$ ($p_{\omega \pi} = 1446$ MeV/$c$ as compared with $p_{K^*
\bar K} = 1373$ MeV/$c$ in $J/\psi \to K^{*+} K^-$ and 1371 MeV/$c$ in $J/\psi
\to K^{*0} \bar K^0$).  For a P-wave decay, the correction factor $\rho^{1/2} =
(p_{K^* \bar K}/p_{\omega \pi})^{3/2} = 0.925$ should thus multiply the square
root of the $\omega \pi^0$ branching ratio in extracting the amplitude
satisfying (\ref{eqn:VPtri}). We then find 
$$
|{\cal A}(J/\psi \to K^{*+} K^-)| = 0.050 \pm 0.002~~,~~~
|{\cal A}(J/\psi \to K^{*0} \bar K^0)| = 0.046 \pm 0.002~~,~~~
$$
\beq
|{\cal A}(J/\psi \to \omega \pi^0)| = |B| = 0.0190 \pm 0.0014~~~.
\eeq
These form a triangle which is roughly isosceles in shape as a result of the
near-equality of the $K^* \bar K$ amplitudes.  The base (the $\omega \pi^0$
side) corresponds to the electromagnetic amplitude $B$, while the sides are
dominated by the strong contribution $A$.

In order to specify the relative phase of strong and electromagnetic
contributions, one needs the ratio $r$.  We take $r \simeq 2/3$ (the
corresponding ratio for constituent-quark masses, which fits electromagnetic
transitions of the form $V \to P \gamma$ \cite{GR}).  The electromagnetic
contributions to ${\cal A}(J/\psi \to K^{*+} K^-)$ and ${\cal A}(J/\psi \to
K^{*0} \bar K^0)$ are then $(4/9)B$ and $-(5/9)B$, respectively, so that the
triangle has the shape illustrated in Fig.~1.  The magnitude of the strong
amplitude is $|A| \simeq 0.047$.

\begin{figure}
\centerline{\epsfysize = 2 in \epsffile {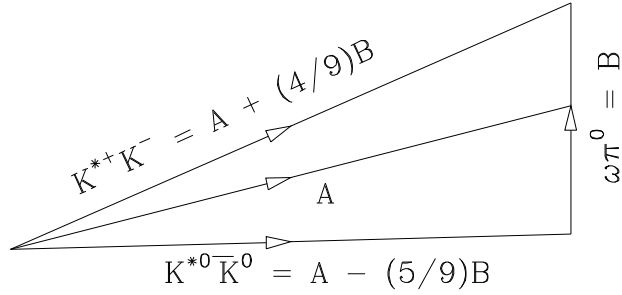}}
\caption{Triangle of amplitudes in $J/\psi \to VP$ decays.  Here $A$ and
$B$ are strong and electromagnetic amplitudes, respectively.}
\end{figure}

A brief calculation of the relative phase $\delta$ of $A$ and $B$ yields the
relation 
\beq
\cos \delta = \frac{{\cal B}(K^{*+} K^-) - {\cal B}(K^{*0} \bar K^0) + \rho
{\cal B}(\omega \pi^0)/9}{2|A||B|} = 0.25 \pm 0.16,
\eeq
or $\delta = (76^{+9}_{-10})^\circ$.  The relative phase between the strong and
electromagnetic amplitudes is large and consistent with $90^\circ$. We have not
made use of the $J/\psi \to \rho \pi$ amplitude since its strong contribution
(which predominates over a very small electromagnetic one) is related to $A$
only through flavor SU(3), which we do not employ.  For similar reasons, we do
not consider other final states such as $\omega \eta$ and $\omega \eta'$.  When
these decays are included in the fit, the results do not change much; Suzuki
\cite{SuzVP} obtains $\delta = 80^\circ$.

A similar analysis yields a large relative phase between strong and
electromagnetic contributions to $J/\psi \to PP$ decays.  We consider the
processes $J/\psi \to K^+ K^-$ and $J/\psi \to K^0 \bar K^0$ [whose strong
amplitudes vanish in the limit of flavor SU(3)] and $J/\psi \to \pi^+ \pi^-$
[whose strong amplitude vanishes in the limit of isospin conservation].
The corresponding amplitudes may be expressed as
$$
{\cal A}(J/\psi \to K^+ K^-) = A' + B'(Q_u - Q_s)~~,~~~
{\cal A}(J/\psi \to K^0 \bar K^0) = A' + B'(Q_d - Q_s)~~,~~~
$$
\beq
{\cal A}(J/\psi \to \pi^+ \pi^-) = A' + B'(Q_u - Q_d)~~~,
\eeq
satisfying the triangle relation
\beq \label{eqn:PPtri}
{\cal A}(J/\psi \to K^+ K^-) - {\cal A}(J/\psi \to K^0 \bar K^0) =
{\cal A}(J/\psi \to \pi^+ \pi^-)~~~.
\eeq
This relation is violated slightly by some of the SU(3)-breaking terms
considered by Suzuki \cite{SuzPP}, but is sufficient for our purposes. As in
the previous calculation, we correct the $\pi^+ \pi^-$ amplitude for the
center-of-mass 3-momentum in $J/\psi \to \pi^+ \pi^-$ ($p_\pi = 1542$ MeV/$c$
as compared with $p_K = 1468$ MeV/$c$ in $J/\psi \to K^+ K^-$ and 1466 MeV/$c$
in $J/\psi \to K^0 \bar K^0$).  The P-wave correction factor to the square root
of the $\pi^+ \pi^-$ branching ratio, needed to extract the amplitude
satisfying (\ref{eqn:PPtri}), is then $\rho^{1/2} = (p_K/p_\pi)^{3/2} = 0.929$.

We use the branching ratios \cite{PDG}
$$
{\cal B}(J/\psi \to K^+ K^-) = (2.37 \pm 0.31) \times 10^{-4}~~,~~~
{\cal B}(J/\psi \to K^0 \bar K^0) = (1.08 \pm 0.14)\times 10^{-4}~~,~~~
$$
\beq
{\cal B}(J/\psi \to \pi^+ \pi^-) = (1.47 \pm 0.23)\times 10^{-4}~~~.
\eeq
The amplitudes which are the square roots of these branching ratios, correcting
the $\pi^+ \pi^-$ amplitude for the kinematic factor mentioned above, are
$$
|{\cal A}(J/\psi \to K^+ K^-)| = |A' + B'| = (1.54 \pm 0.10) \times 10^{-2}~~~,
$$
$$
|{\cal A}(J/\psi \to K^0 \bar K^0)| = |A'| = (1.04 \pm 0.07) \times 10^{-2}~~~,
$$
\beq
|{\cal A}(J/\psi \to \pi^+ \pi^-)| = |B'| = (1.12 \pm 0.08) \times 10^{-2}~~~.
\eeq
These form an isosceles right triangle, as illustrated in Fig.~2. 

\begin{figure}
\centerline{\epsfysize = 2.7 in \epsffile {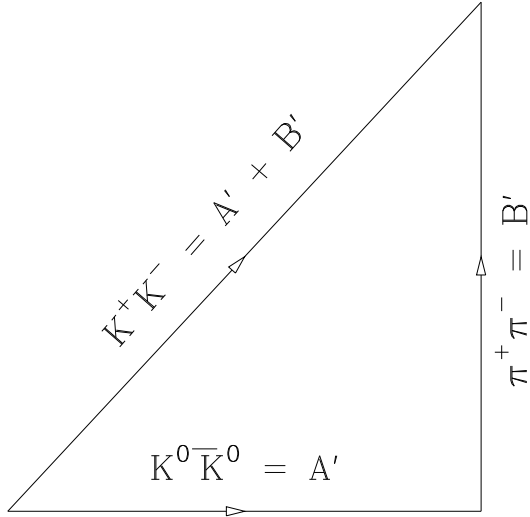}}
\caption{Triangle of amplitudes in $J/\psi \to PP$ decays.  Here $A'$ and
$B'$ are strong and electromagnetic amplitudes, respectively.}
\end{figure}

The relative phase $\delta'$ of $A'$ and $B'$ is given by
\beq
\cos \delta' = \frac{{\cal B}(K^+ K^-) - {\cal B}(K^0 \bar K^0) -
\rho {\cal B}(\pi^+ \pi^-)}{2|A'||B'|} = 0.01 \pm 0.19,
\eeq
or $\delta' = (89 \pm 10)^\circ$.  This is again in accord with Suzuki's result
\cite{SuzPP}.
\bigskip

\leftline{\bf B.  $\psi'$ decays}
\bigskip

The suppression of certain decay modes of the $\psi'$ such as $\rho \pi$ and
$K^* \bar K + {\rm~c.c.}$ \cite{MEBF} has puzzled physicists for nearly 20
years \cite{rhopi}.  The BES Collaboration \cite{BESDPF} has now reported the
isospin-violating decay $\psi' \to \omega \pi^0$ at a level above the upper
limit for the isospin-allowed decay $\psi' \to \rho \pi$, and has seen the
decay $\psi' \to K^{*0} \bar K^0$ at a level considerably above the upper
limit for the isospin-related decay $\psi' \to K^{*+} K^-$. These results are
summarized in Table I, whose data are taken from Ref.~\cite{BESDPF}.

An analysis similar to that performed for $J/\psi \to VP$ yields the amplitudes
(expressed again as square roots of branching ratios, with a kinematic
correction for $\omega \pi^0$)
$$
|{\cal A}(\psi' \to K^{*+} K^-)| < 3.9 \times 10^{-3}~~,~~~
|{\cal A}(\psi' \to K^{*0} \bar K^0)| = (6.4 \pm 2.3) \times 10^{-3}~~,~~~
$$
\beq
|{\cal A}(\psi' \to \omega \pi^0)| = (5.8 \pm 1.6) \times 10^{-3}~~~.
\eeq
These should satisfy the sum rule (\ref{eqn:VPtri}) with $\psi'$ replacing
$J/\psi$.

% This is Table I
\begin{table}
\caption{Branching ratios of the $\psi'$ to specific hadronic final states.}
\begin{center}
\begin{tabular}{c c} \hline \hline
Final & B.r. or 90\% c.l. \\
state & upper limit ($\times 10^{-5}$)\\ \hline
$\rho \pi$ &  $< 2.8$ \\
$\omega \pi^0$ & $3.8 \pm 1.7 \pm 1.1$ \\
$K^{*+} K^-$ + c.c. & $< 3.0$ \\
$K^{*0} \bar K^0$ + c.c. & $8.1 \pm 2.4 \pm 1.6$ \\ \hline \hline
\end{tabular}
\end{center}
\end{table}

The data are not yet precise enough to specify the shape of the corresponding
$\psi' \to VP$ amplitude triangle.  The $\omega \pi^0$ decay requires
an electromagnetic contribution to be present.  If this were
the only amplitude contributing to all three processes, one would expect ${\cal
A}(K^{*+} K^-) = (4/5) {\cal A}(K^{*0} \bar K^0) = (4/9) {\cal A}(\omega
\pi^0)$, which is just at the limit of error bars for each amplitude, but not
yet firmly ruled out.  (Take, for example, ${\cal A}(\omega \pi^0) = 7.4 \times
10^{-3},$ ${\cal A}(K^{*0} \bar K^0) = 4.1 \times 10^{-3}$, and ${\cal
A}(K^{*+} K^-) = 3.3 \times 10^{-3}$.) Thus, since the presence of the strong
amplitude has not yet been demonstrated, its phase with respect to the
electromagnetic one is still an open question.
\bigskip

\centerline{\bf III.  CHARMED MESON DECAYS}
\bigskip

\leftline{\bf A.  Isospin decomposition}
\bigskip

The decays of the nonstrange charmed mesons $D^+ = c \bar d$ and $D^0 = c \bar
u$ to final states consisting of one strange meson ($\bar K$ or $\bar K^*$) and
one $I=1$ nonstrange meson ($\pi$ or $\rho$) are governed by the $\Delta I =
\Delta I_3 = 1$ subprocess $c \to s u \bar d$ and thus are characterized by two
amplitudes $A_{1/2}$ and $A_{3/2}$ labeled by the total isospin of the final
state.  For example, the amplitudes for the decays $D \to \bar K \pi$ are given
by 
$$
{\cal A}(D^+ \to \bar K^0 \pi^+) = A_{3/2}~~,~~~
{\cal A}(D^0 \to K^- \pi^+) = \frac{2}{3}A_{1/2} + \frac{1}{3}A_{3/2}~~~,
$$
\beq
{\cal A}(D^0 \to \bar K^0 \pi^0) = -\frac{\s}{3}A_{1/2} + \frac{\s}{3}
A_{3/2}~~~,
\eeq
and thus satisfy a triangle relation
\beq \label{eqn:Dtri}
{\cal A}(\bar K^0 \pi^+) = {\cal A}(K^- \pi^+) + \s {\cal A}(\bar K^0 \pi^0)
~~~,
\eeq
where we have omitted the initial particle. By studying decay rates alone, one
can determine the shape of this triangle and thus learn the relative phases of
isospin amplitudes.  We shall continue the discussion with the $D \to \bar K
\pi$ example; it also holds for $D \to \bar K^* \pi$ and $D \to \bar K \rho$.
We do not use information obtained in some analyses
\cite{MkIII,Anjos,ARGUS,Frab} from relative phases of bands in Dalitz plots,
but will return to this question in the subsequent discussion. 

The magnitude of the $I = 3/2$ amplitude is obtained from the $D^+ \to \bar K^0
\pi^+$ partial width.  Omitting all kinematic factors, we have $|A_{3/2}|^2 =
\Gamma(\bar K^0 \pi^+)$.  The magnitude of the $I = 1/2$ amplitude is obtained
from the combination 
\beq
|A_{1/2}|^2 = \frac{3}{2} \left[ \Gamma(K^- \pi^+) + \Gamma(\bar K^0 \pi^0)
\right] - \frac{1}{2} \Gamma(\bar K^0 \pi^+)~~~.
\eeq
The relative phase $\delta_I$ between isospin amplitudes is given by
\beq
\cos \delta_I = \frac{3 \Gamma(K^- \pi^+) + \Gamma(\bar K^0 \pi^+)
- 6 \Gamma(\bar K^0 \pi^0)} {4|A_{1/2} A_{3/2}|}~~~.
\eeq
\bigskip

\leftline{\bf B.  Graphical decomposition}
\bigskip

As stressed by Suzuki \cite{SuzDB}, for decays of $D$ and $B$ mesons in which
multi-particle final states can play a large role, two-body isospin amplitudes
may not be the most significant quantities. The relation (\ref{eqn:Dtri}) also
is implied by the decomposition of the decay amplitudes in terms of
color-favored tree ($T$), color-suppressed ($C$), and exchange ($E$) amplitudes
\cite{ZepSWChau,GHLR}: 
\beq
{\cal A}(\bar K^0 \pi^+) = T + C~~,~~~ {\cal A}(K^- \pi^+) = T + E~~,~~~ {\cal
A}(\bar K^0 \pi^0) = (C - E)/\s~~~.
\eeq 
The set $T$, $C$, and $E$ is over-complete.  In principle one can assume that
$T$ and $C$ have zero phase relative to one another, and that all the
final-state interaction effects are concentrated in $E$.  One still needs
information on the relative magnitude of $T$ and $C$, which one may either take
from QCD \cite{BSW,NS}, or by applying the factorization hypothesis and the
relation between nonleptonic and semileptonic processes \cite{BJ,JLRFM}, 
\beq \label{eqn:BJ}
\frac{\Gamma(D^0 \to K^- \pi^+)_T}{d\Gamma(D^0 \to K^- \ell^+ \nu_\ell)/dq^2|
_{q^2 = m_\pi^2}} = 6 \pi^2 f_\pi^2 |V_{ud}|^2 = 0.98 {\rm~GeV}^2~~~,
\eeq
to data on the spectrum in semileptonic decays (see, e.g., \cite{E687spec}).
The use of a phenomenological $E$ amplitude to parametrize final-state
interactions would be an alternative to the more conventional short-distance
descriptions of such effects in charmed-particle decays, which can account
for some but perhaps not all of the differences among charmed-particle
lifetimes \cite{Shifman}. We leave this possibility for a future investigation.
\bigskip

\leftline{\bf C.  Results of isospin analysis}
\bigskip

To compare amplitudes for $D^0$ and $D^+$ decays, we calculate decay rates
using branching ratios and lifetimes.  The Particle Data Group values
\cite{PDG} are averaged with new CLEO values \cite{CLEOlife} in Table II; we
use the new averages in what follows.  We summarize the relevant branching
ratios and decay rates in Table III. 

The values of isospin amplitudes (defined as square roots of rates, without
any correction for kinematic factors) and corresponding phases are shown in
Table IV.  In accord with many previous results \cite{MkIII,Anjos,ARGUS,Frab},
the relative phases of the $I = 1/2$ and $I = 3/2$ amplitudes are consistent
with $90^\circ$ for the $\bar K \pi$ and $\bar K^* \pi$ channels and with
$0^\circ$ for the $\bar K \rho$ channel.  The value of $\delta_I(\bar K \pi)$
agrees with that of Suzuki \cite{SuzDB}.  The relation of the amplitudes to one
another is illustrated in Fig.~3, showing the relative phase near $90^\circ$. 

% This is Table II
\begin{table}
\caption{Charmed meson lifetimes, in fs.}
\begin{center}
\begin{tabular}{c c c c} \hline \hline
State & PDG & CLEO & Average \\ \hline
$D^+$ & $1057 \pm 15$ & $1033.6 \pm 22.1~^{+9.9}_{-12.7}$ & $1051 \pm 13$ \\
$D^0$ & $415 \pm 4$ & $408.5 \pm 4.1~^{+3.5}_{-3.4}$ & $412.7 \pm 3.2$ \\
\hline \hline
\end{tabular}
\end{center}
\end{table}

% This is Table III
\begin{table}
\caption{$D^+$ and $D^0$ ranching ratios and decay rates.}
\begin{center}
\begin{tabular}{c c c} \hline \hline
Mode & Branching ratio & Decay rate \\
     & (percent)       & ($\times 10^{10} {\rm s}^{-1}$) \\ \hline
\multicolumn{3}{c}{$D^+$ decays} \\ \hline
$\bar K^0 \pi^+$    & $2.89 \pm 0.26$ & $2.75 \pm 0.25$ \\
$\bar K^{*0} \pi^+$ & $1.90 \pm 0.19$ & $1.81 \pm 0.18$ \\
$\bar K^0 \rho^+$   & $6.6 \pm 2.5$   & $6.3 \pm 2.4$   \\ \hline
\multicolumn{3}{c}{$D^0 \to (-+)$ decays} \\ \hline
$K^- \pi^+$         & $3.85 \pm 0.09$ & $9.33 \pm 0.23$ \\
$K^{*-} \pi^+$      &  $5.1 \pm 0.4$  & $12.4 \pm 1.0$  \\
$K^- \rho^+$        & $10.8 \pm 1.0$  & $26.2 \pm 2.4$  \\ \hline
\multicolumn{3}{c}{$D^0 \to (00)$ decays} \\ \hline
$\bar K^0 \pi^0$    & $2.12 \pm 0.21$ & $5.14 \pm 0.51$ \\
$\bar K^{*0} \pi^0$ &  $3.2 \pm 0.4$  &  $7.8 \pm 1.0$  \\
$\bar K^0 \rho^0$   & $1.21 \pm 0.17$ & $2.93 \pm 0.41$ \\ \hline \hline
\end{tabular}
\end{center}
\end{table}

The amplitude triangle for $D \to K^* \pi$ (Fig.~4) is qualitatively similar
to that in Fig.~3, but the $(00)$ and $(-+)$ sides are longer in proportion to
the $(0+)$ side, so $|A_{1/2}/A_{3/2}|$ is larger.  The amplitude relation for
$D \to \bar K \rho$ degenerates into a straight line since the central value of
$\cos \delta_I$ exceeds 1.  Put differently, the square roots of the $\bar K
\rho$ rates in Table III satisfy $(K^- \rho^+)^{1/2} > (\bar K^0 \rho^+)^{1/2}
+ (2 \bar K^0 \rho^0)^{1/2}$.

% This is Table IV
\begin{table}
\caption{Isospin amplitudes and relative phases $\delta_I$ for $D$ decays.}
\begin{center}
\begin{tabular}{c c c c} \hline \hline
Mode & $|A_{1/2}|$ & $|A_{3/2}|$ & $\delta_I$ \\
     & \multicolumn{2}{c}{$(\times 10^5 {\rm s}^{-1/2}$)} & (degrees) \\ \hline
$\bar K \pi$   & $4.51 \pm 0.23$ & $1.66 \pm 0.08$ & $90 \pm 7$   \\
$\bar K^* \pi$ & $5.43 \pm 0.19$ & $1.35 \pm 0.07$ & $105 \pm 14$ \\
$\bar K \rho$  & $6.36 \pm 0.30$ & $2.51 \pm 0.47$ & $ < 27~(1 \sigma)$ \\
\hline \hline
\end{tabular}
\end{center}
\end{table}

\begin{figure}
\centerline{\epsfysize = 2 in \epsffile {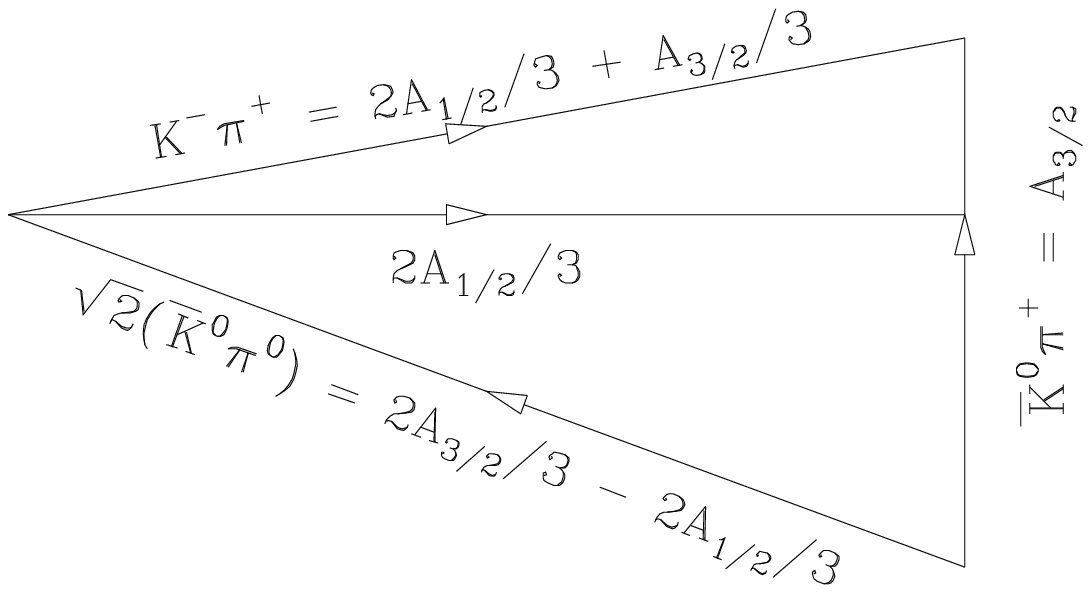}}
\caption{Triangle of amplitudes in $D \to \bar K \pi$ decays.  Subscripts on
amplitudes denote total isospin.}
\end{figure}

\begin{figure}
\centerline{\epsfysize = 2 in \epsffile {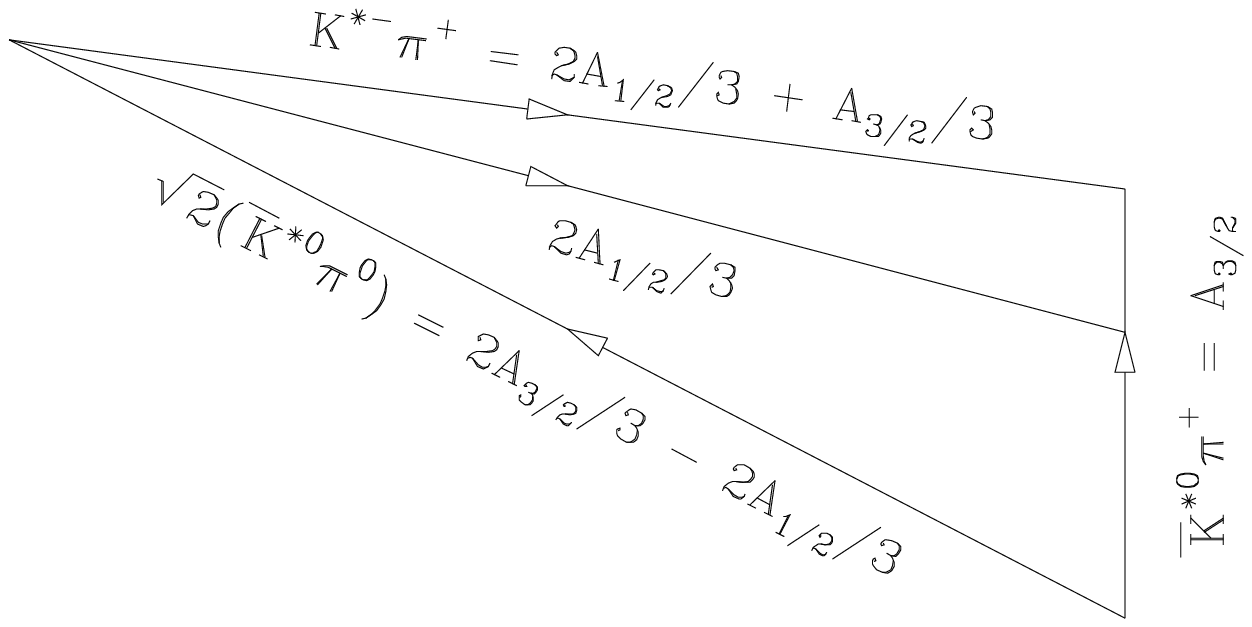}}
\caption{Triangle of amplitudes in $D \to \bar K^* \pi$ decays.  Subscripts on
amplitudes denote total isospin.}
\end{figure}

Resonances in $I = 1/2$ channels (they have never been seen in $I = 3/2$
channels) can contribute to the $D \to \bar K \pi$ and $\bar K^* \pi$ processes
\cite{HJL,Kamal,Banff}.  One needs two different states since the $\bar K \pi$
state with total angular momentum $J = 0$ has even parity, while the $J = 0$
$\bar K^* \pi$ state has odd parity.  Candidates for the even-parity
\cite{Aston} and odd-parity \cite{Armstrong} state exist. The odd-parity
resonance should couple much more strongly to $K^* \pi$ than to $\bar K \rho$
in order to explain the absence of a large final-state phase in the $\bar K
\rho$ channel.  However, it has only been reported in the $K \phi$ channel
\cite{Armstrong}. 
\bigskip

\leftline{\bf D.  Interference between bands on Dalitz plot}
\bigskip

In Dalitz plot analyses of $D \to \bar K^* \pi$ and $D \to \bar K \rho$,
several cross-checks of relative phases of amplitudes can be performed
\cite{MkIII,Anjos,ARGUS,Frab}.  We enumerate each three-body final state and
the information it provides. 

1)  $D^0 \to K^- \pi^+ \pi^0$ contributes to $K^- \rho^+$, $K^{*-} \pi^+$, and
$\bar K^{*0} \pi^0$.  The amplitude triangle construction for $D \to \bar K^*
\pi$ implies a relative phase between the $K^{*-} \pi^+$ and $\bar K^{*0}
\pi^0$ amplitudes (cf. Fig.~4) of 
\beq \label{eqn:Ks1}
\delta_{K^{*-} \pi^+, \bar K^{*0} \pi^0} = \cos^{-1} \frac{\Gamma(\bar K^{*0}
\pi^+) - 2 \Gamma(\bar K^{*0} \pi^0) - \Gamma(K^{*-} \pi^+)}{2 \s |{\cal A}
(\bar K^{*0} \pi^0) {\cal A}(K^{*-} \pi^+)|} = (160^{+20}_{-14})^\circ~~~.
\eeq
The E687 \cn \cite{Frab}, for comparison, obtains $\delta_{K^{*-} \pi^+, K^-
\rho^+} = (162 \pm 10 \pm 7 \pm 4)^\circ$ and $\delta_{\bar K^{*0} \pi^0, K^-
\rho^+} = (-2 \pm 12 \pm 23 \pm 2)^\circ$, while the Mark III \cn~\cite{MkIII}
finds $\delta_{K^{*-} \pi^+, K^- \rho^+} = (154 \pm 11)^\circ$ and
$\delta_{\bar K^{*0} \pi^0, K^- \rho^+} = (7 \pm 7)^\circ$. In both cases the
$K^- \rho^+$ amplitude was taken to be real in the analysis of the $K^- \pi^+
\pi^0$ final state.  The first two E687 errors are statistical and systematic,
respectively.  The last E687 error is associated with the uncertainty in the
relative contributions of specific final states to the Dalitz plot.  The
agreement with (\ref{eqn:Ks1}) is satisfactory. 

The E691 \cn~chooses a reference phase of $0^\circ$ for the nonresonant
amplitude.  With respect to this phase, they find  $\delta_{K^{*-} \pi^+} =
(-112 \pm 9)^\circ$, $\delta_{\bar K^{*0} \pi^0} = (167 \pm 9)^\circ$,
$\delta_{K^- \rho^+} = (40 \pm 7)^\circ$.  It is less clear whether this
result agrees so well with (\ref{eqn:Ks1}).

2) $D^0 \to \bar K^0 \pi^+ \pi^-$ contributes to $K^{*-} \pi^+$ and $\bar K^0
\rho^0$.  Our previous discussion implies that $\bar K^0 \rho^0$ and $K^-
\rho^+$ should be relatively real, so we expect $\delta_{K^{*-} \pi^+, \bar K^0
\rho^0} = \delta_{K^{*-} \pi^+, K^- \rho^+}$.  Ref.~\cite{Frab} obtains $(136
\pm 6 \pm 2 \pm 2)^\circ$ while Ref.~\cite{ARGUS} obtains $(137 \pm 7)^\circ$
for the left-hand side, in adequate but not perfect agreement with the value
quoted above for the right-hand side. Ref.~\cite{MkIII} obtains $\delta_{\bar
K^0 \rho^0} = (93 \pm 30)^\circ$ in a convention in which $\delta_{K^{*-}
\pi^+} = 0$.  This is not particularly close to (\ref{eqn:Ks1}). 
Ref.~\cite{Anjos} finds phases of $\Delta_{K^{*-} \pi^+} = (109 \pm 9)^\circ$
and $\Delta_{\bar K^0 \rho^0} = (-123 \pm 12)^\circ$ with respect to the
nonresonant amplitude. 

3) $D^0 \to \bar K^0 \pi^0 \pi^0$ has two identical $\bar K^{*0} \pi^0$ bands
which should interfere constructively with one another. 

4) $D^+ \to \bar K^0 \pi^+ \pi^0$ contributes to $\bar K^{*0} \pi^+$ and $\bar
K^0 \rho^+$.  The amplitude triangles predict that (a) all the $\bar
K \rho $ amplitudes are relatively real, and (b) the relative phase between the
$\bar K^{*0} \pi^+$ and $K^{*-} \pi^+$ amplitudes (cf. Fig.~4) is
\beq \label{eqn:Ks2}
\delta_{\bar K^{*0} \pi^+, K^{*-} \pi^+} = \cos^{-1} \frac{\Gamma(\bar K^{*0}
\pi^+) + \Gamma(\bar K^- \pi^+) - 2 \Gamma(\bar K^{*0} \pi^0)}{2 |{\cal A}
(\bar K^{*0} \pi^+) {\cal A}(K^{*-} \pi^+)|} = (98^{+14}_{-13})^\circ~~~.
\eeq
These are tested by Mark III data on the $D^+ \to \bar K^0 \pi^+ \pi^0$ final
state.  One must combine the results $\delta_{\bar K^{*0} \pi^+, \bar K^0
\rho^+} = (43 \pm 23)^\circ$ from this final state with the previously
mentioned phase $\delta_{K^{*-} \pi^+, K^- \rho^+} = (154 \pm 11)^\circ$ from
this experiment; the agreement seems good. 

5) $D^+ \to K^- \pi^+ \pi^+$ has two identical $\bar K^{*0} \pi^+$ bands which
should interfere constructively with one another.

Ref.~\cite{Frab} contains some comments on the possibility that not all
experiments quote phases with the same convention.
\bigskip
% \newpage

\centerline{\bf IV.  PENGUIN-DOMINATED $b \to s$ PROCESSES}
\bigskip

\leftline{\bf A.  Charm-anticharm annihilation}
\bigskip

A number of features of $B$ decays suggest a possible role for enhanced
charm-anticharm annihilation into non-charmed final states \cite{Dunietz}:

\begin{enumerate}

\item The semileptonic branching ratio $\b(B \to X \ell \nu)$ is about 11\%
(vs.~a theoretical prediction of about 12\%) \cite{MN}.

\item The number $n_c$ of charmed particles per average $B$ decay is about 1.1
to 1.2 vs.~a theoretical prediction of 1.2 to 1.3 \cite{MN}. 

\item The inclusive branching ratio $\b(B \to \eta' X)$ appears large
\cite{CLEOeta} in comparison with theoretical expectations \cite{incl}.

\item The exclusive branching ratio $\b(B \to K \eta')$ \cite{CLEOeta} appears
to require an additional contribution \cite{eta} in comparison with the
penguin contribution leading to $B^0 \to K^+ \pi^-$ or $B^+ \to K^0 \pi^+$. 

\end{enumerate}

A common source for these effects could be an enhanced rate for the subprocess
$\bar b \to \bar c c s \to \bar q q \bar s$, where $q$ stands for a light
quark, e.g., through rescattering effects.  These are inherently long-range and
nonperturbative and could also be responsible for the overall enhancement of
the $\bar b \to \bar s$ penguin transitions noted in Refs.~\cite{Ciu}.
Alternatives for points (3) and (4) which have been suggested include a large
$c \bar c$ \cite{KBe} or gluonic component in the $\eta'$. The former
possibility is intriguing but one must then ascribe the suppression of the
decay $J/\psi \to \eta' \gamma$ to form factor effects. 

If rescattering from the $\bar b \to \bar c c \bar s$ subprocess into states
containing light quarks really is important, both the overall $\bar b \to \bar
s$ penguin amplitude and a specific contribution \cite{eta} to $\bar b \to \bar
s + (\eta, \eta')$ (to be mentioned below) could have strong phases very
different from the tree amplitude contributing to $B \to K + X$ decays, raising
the possibility of substantial CP-violating asymmetries whenever these
amplitudes interfere with one another in a self-tagging $B$ decay (such as $B^0
\to K^+ \pi^-$).  We shall now indicate where such effects are likely to be
visible.  (See also Refs.~\cite{GRcomb,Neu,DHHP,HHY}.) 
\bigskip

\leftline{\bf B.  Estimate of amplitudes and application to decays involving
$\eta'$} 
\bigskip

In what follows we shall update an estimate \cite{DGRPRL} of the amplitudes
contributing to the decays of $B \to PP$, where $P$ is a light pseudoscalar
meson.  These amplitudes are denoted by $t$ (tree), $p$ (penguin), and $s$
(singlet penguin).  Color-suppressed amplitudes and electroweak penguin
amplitudes \cite{EWP,GHLREWP,NR,GPY} are neglected for simplicity.  We shall be
concerned with the relative strong phases of these amplitudes, which if large
could lead to observable CP-violating asymmetries in several interesting final
states. Amplitudes for strangeness-preserving processes will be unprimed, while
those for strangeness-changing processes will be primed. 

The weak phases for strangeness-preserving processes are ${\rm arg}(t) = {\rm
arg}(V_{ub}^* V_{ud}) = \gamma$ and ${\rm arg}(p) = {\rm arg}(s) \simeq {\rm
arg}(V_{tb}^* V_{td}) = - \beta$, so that the relative phase of $t$ and $p$ or
$s$ amplitudes is $\gamma + \beta = \pi - \alpha$.  Here $\alpha$, $\beta$, and
$\gamma$ are angles of the unitarity triangle as defined, for example, in
Ref.~\cite{NQ}.  (They are also referred to as $\phi_2$, $\phi_1$, and
$\phi_3$, respectively \cite{HRS}.) For strangeness-changing processes the
expected phases are ${\rm arg}(t') = {\rm arg}(V_{ub}^* V_{us}) = \gamma$ and
${\rm arg}(p') = {\rm arg}(s') \simeq {\rm arg}(V_{tb}^* V_{ts}) = \pi$.  Thus,
the relative phase of $t'$ and $p'$ or $s'$ amplitudes is $\gamma$ (modulo
$\pi$).

The tree amplitude $t$ is expected to dominate strangeness-preserving $B \to
PP$ decays such as $B^+ \to \pi^+ \pi^0$ and $B^0 \to \pi^+ \pi^-$.  Although
no conclusive evidence has been presented for these decays, one estimates
\cite{GRcomb} using factorization and the semileptonic process $B \to \pi \ell
\nu$ that ${\cal B}(B^+ \to \pi^+ \pi^0) \simeq (1/2) {\cal B} (B^0 \to \pi^+
\pi^-) \simeq 4 \times 10^{-6}$. One can then use the relation $t' \simeq
\lambda t$ to estimate the magnitude of the tree amplitude in
strangeness-changing processes.  Here $\lambda \simeq 0.2$ is the parameter
introduced by Wolfenstein \cite{WP} to describe the hierarchy of CKM elements. 

The penguin amplitude $p'$ is expected to dominate strangeness-changing $B \to
PP$ decays such as $B^0 \to K^+ \pi^-$ and in particular $B^+ \to K^0 \pi^+$
(which has no $t'$ contribution).  Differences between ${\cal B}(B^0 \to K^+
\pi^-)$, ${\cal B}(B^+ \to K^0 \pi^+)$, and $2 {\cal B}(B^+ \to K^+ \pi^0)$,
important in more precise treatments which include effects of interference on
rates \cite{GRcomb,NR}, will be ignored here.

The coefficients of amplitudes in each decay process are given in
Ref.~\cite{DGRPRL}.  Using the most recent rates for $B \to PP$ decays
\cite{GRcomb,CLEOPP}, we find the results shown in Table V.  These deserve
several comments. 

% This is Table V
\begin{table}
\caption{Summary of predicted contributions to selected $\Delta S = 0$ decays
of $B$ mesons.  Rates are quoted in branching ratio units of $10^{-6}$.  Rates
in italics are assumed inputs.} 
\begin{center}
\begin{tabular}{r c c c c} \hline \hline
$\Delta S = 0$ & $|t|^2$ & $|p|^2$ & \multicolumn{2}{c} {$|s|^2$ rate} \\
Decay       &   rate  &   rate  &   (a)    & (b) \\ \hline
$B^+ \to \pi^+ \pi^0$ & {\it 4} & 0 & 0 & 0 \\
$ \to \pi^+ \eta$ & 2.7 & 1.0 & 0.09 & 0.3 \\
$ \to \pi^+ \eta'$ & 1.3 & 0.5 & 0.7 & 2.4 \\
\hline
$B^0 \to \pi^+ \pi^-$ & {\it 8} & 0.7 & 0 & 0 \\ \hline \hline
$|\Delta S = 1|$ & $|t'|^2$ & $|p'|^2$ & \multicolumn{2}{c} {$|s'|^2$ rate} \\
Decay       &   rate   &   rate   &  (a) & (b) \\ \hline
$B^+ \to K^0 \pi^+$ & 0 & {\it 14} & 0 & 0 \\
$ \to K^+ \pi^0$ & 0.2 & 7& 0 & 0 \\
$ \to K^+ \eta$ & 0.13 & $\simeq 0$ & 1.7 & 6 \\
$ \to K^+ \eta'$ & 0.07 & 21 & 14 & 48 \\
\hline
$B^0 \to K^+ \pi^-$ & 0.4 & {\it 14} & 0 & 0 \\
$ \to K^0 \pi^0$ & 0 & 7 & 0 & 0 \\
$ \to K^0 \eta$ & 0 & $\simeq 0$ & 1.7 & 6 \\
$ \to K^0 \eta'$ & 0 & 21 & 14 & 48 \\ \hline \hline
\end{tabular}
\end{center}

\leftline{(a): Constructive interference between $p'$ and $s'$ amplitudes
assumed in $B^+ \to K^+ \eta'$.}
\leftline{(b): No interference between $p'$ and $s'$ amplitudes
assumed in $B^+ \to K^+ \eta'$.}

\end{table}

1) The $s'$ amplitude is needed in order to properly describe the large rate
\cite{CLEOeta} ${\cal B}(B \to K \eta') = (69 \pm 12) \times 10^{-6}$.  Here we
have averaged the values quoted for charged and neutral $B$ decays.  If the
$s'$ amplitude interferes constructively with $p'$, it does not have to be as
large in magnitude as $p'$, as one sees by comparing the $|p'|^2$ rate for $B
\to K \eta'$ in Table V with the $|s'|^2$ rate from column (a) of the same
table.  The weak phases of $s'$ and $p'$ are expected to be the same, aside
from possible small electroweak penguin effects \cite{GHLREWP}.  The strong
strong phases of these two amplitudes could well be equal as well if they are
both dominated by a large imaginary part associated with the annihilation of a
$c \bar c$ pair into light quarks.  Such a predominantly imaginary amplitude is
one possible interpretation of the large final-state phases \cite{SuzVP,SuzPP}
in certain $J/\psi$ hadronic decays which were discussed in Sec.~II. 

2) The possibility for large CP-violating asymmetries exists whenever two
weak amplitudes $a_1$ and $a_2$ [cf. Eq.~(3)] are not too dissimilar in
magnitude and the sines of both their weak phase difference $\phi_1 - \phi_2$
and their strong phase difference $\delta_1 - \delta_2$ are close to 1.  In
Table VI we identify a few of these interesting cases.

% This is Table VI
\begin{table}
\caption{Examples of possible direct CP asymmetries in $B$ decays}
\begin{center}
\begin{tabular}{c c c c} \hline \hline
Process & Interfering &  Relative  & Maximum   \\
        & amplitudes  & weak phase & asymmetry \\ \hline
$B^+ \to \pi^+ \eta$  & $t,p$ & $\pi - \alpha$ & $\sqrt{3/4}$ \\
$B^+ \to \pi^+ \eta'$ & $t,s$ & $\pi - \alpha$ & 1 \\
$B^+ \to K^+ \pi^-$  & $p',t'$ & $\gamma$ & 0.34 \\
$B^0 \to K^+ \pi^-$ & $p',t'$ & $\gamma$ & 0.34 \\ \hline \hline
\end{tabular}
\end{center}
\end{table}

Although the $t'$ amplitude in $B^+ \to K^+ \pi^0$ and $B^0 \to K^+ \pi^-$
processes is expected to be considerably smaller than the dominant $p'$
amplitude, it can have a noticeable effect on the asymmetry if the strong phase
difference is large.  When the asymmetries in these two processes are combined,
one may even be able to see an effect with present or modestly improved
statistics \cite{GRcomb}.  The asymmetries in $B^+ \to K^+ \pi^0$ and $B^0 \to
K^+ \pi^-$ are expected to be highly correlated \cite{GRcomb,Neu}.

To summarize, we are suggesting the prospect of a large strong phase shift
difference $\delta_1 - \delta_2$ in certain two-body decays of $B$ mesons to
pairs of light pseudoscalar mesons, when one of the weak amplitudes ($p, p', s$,
or $s'$) has a large strong phase difference with respect to the other ($t$ or
$t'$).  Such a phase may arise as a result of strong absorptive effects in
rescattering of $c \bar c$ to light quarks.  Although a perturbative
calculation at the quark level \cite{BSS} gives a small final-state phase,
the possibility that it could be larger (even maximal, i.e., near $\pi/2$)
was suggested some time ago \cite{BPT}. The rescattering of $c \bar c$ to light
final states can enhance the $\bar b \to \bar s$ penguin amplitude without
affecting its weak phase, which remains real: ${\rm arg}(V_{cb}^* V_{cs}) = 0$
vs.~${\rm arg}(V_{tb}^* V_{ts}) = \pi$. 
\bigskip

\leftline{\bf C.  Annihilation of light quarks}
\bigskip

If the $\bar b \to \bar s$ penguin amplitude receives important contributions
from the tree subprocess $\bar b \to \bar u u \bar s$, followed by rescattering
to another final state (such as $\bar d d \bar s$), the estimate of the weak
phase of the amplitude for $B^+ \to K^0 \pi^+$ may be called into question
\cite{resc}.  Normally one expects this process to have a weak phase of $\pi$
or zero, so that there should be only a very small CP-violating difference
between the rates for $B^+ \to K^0 \pi^+$ and $B^- \to \bar K^0 \pi^-$.  This
difference, in the absence of rescattering, would be due entirely to the
process in which the $\bar b$ and spectator $u$ in a $B^+$ annihilate one
another through a virtual $W$ which then produces $K^0 \pi^+$.  Such an
amplitude is expected to be suppressed by a factor of $f_B/m_B$ in comparison
with the dominant ones in which the spectator quark does not participate.

The decay $B^0 \to K^+ K^-$ is particularly sensitive to spectator quark
effects since the $B^0$ contains a $d$ quark which is not present in the final
state \cite{GHLR,GHLRresc}.  It must occur through the process $\bar b d \to
\bar u u$, in which the $\bar u u$ pair either fragments into $K^+ K^-$
directly or annihilates into a multi-gluon state which then materializes as
$K^+ K^-$. Alternatively, it can be fed by rescattering from such final states
as $B^0 \to M_1^+ M_2^-$, where $M_i$ are non-strange mesons like $\pi$ and
$\rho$. Thus, a good way to gauge the effects of this rescattering is to
measure the branching ratio for $B^0 \to K^+ K^-$.  If it exceeds the value of
a few times $10^{-8}$, one must take rescattering effects seriously. 

Another method which has been proposed to estimate rescattering effects is to
study the rate and CP-violating asymmetry for the decay $B^+ \to \bar K^0 K^+$,
whose amplitudes are related to those in $B^+ \to K^0 \pi^+$ by flavor SU(3)
\cite{resc}.  Specifically, the penguin amplitude in $B^+ \to \bar K^0 K^+$
should be suppressed by $|V_{td}/V_{ts}| = {\cal O}(\lambda) \simeq 1/5$ with
respect to that in $B^+ \to K^0 \pi^+$, while the corresponding annihilation
(or rescattering) amplitude should be {\it enhanced} by $|V_{ud}/V_{us}| =
1/\lambda$.  Our discussion indicates that both the (suppressed) penguin
amplitude and the (enhanced) rescattering amplitude in $B^+ \to \bar K^0 K^+$
may have the same final-state phase characteristic of a highly absorptive
process, so that a CP-violating difference between $\Gamma(B^+ \to \bar K^0
K^+)$ and $\Gamma(B^- \to K^0 K^-)$ may not be visible even if the rescattering
process is playing an important role.  One then falls back on the proposed
enhancement of the total $B^+ \to \bar K^0 K^+$ decay rate, which would require
a substantial rescattering contribution to be observable, or -- better, in our
opinion -- the observation of the rare process $B^0 \to K^+ K^-$ to indicate
the magnitude of rescattering effects.  
\bigskip

\centerline{\bf V.  $B$ DECAYS TO CHARMED FINAL STATES}
\bigskip

\leftline{\bf A.  Decays to $\bar D \pi$, $\bar D^* \pi$, $\bar D \rho$,
$\bar D^* \rho$}
\bigskip

The pattern of $B$ decays to charmed final states has important differences
with respect to the corresponding pattern for $D$ decays to strange states.
First of all, the relative phase of color-suppressed ($C$) and color-favored
($T$) amplitudes is different from that in charm decays \cite{NS}.  Second, one
cannot evaluate the final-state phases associated with rescattering effects
since these effects seem so small. 

We may perform an isospin decomposition similar to that in Sec.~III for charm
decays by noting that the fundamental $\bar b \to \bar c u \bar d$ subprocess
responsible for $B \to \bar D^{(*)} + X$ decays has $\Delta I = \Delta I_3 =
1$.  Thus, when $X$ is an $I = 1$ meson (e.g., $\pi$ or $\rho$), there will
again be two isospin amplitudes.  For $\bar B \to D \pi$ decays we may then
write 

$$
{\cal A}(B^+ \to \bar D^0 \pi^+) = A_{3/2}~~,~~~
{\cal A}(B^0 \to D^- \pi^+) = \frac{2}{3}A_{1/2} + \frac{1}{3}A_{3/2}~~~,
$$
\beq
{\cal A}(B^0 \to \bar D^0 \pi^0) = -\frac{\s}{3}A_{1/2} + \frac{\s}{3}
A_{3/2}~~~.
\eeq
These amplitudes again satisfy a triangle relation
\beq \label{eqn:Btri}
{\cal A}(\bar D^0 \pi^+) = {\cal A}(D^- \pi^+) + \s {\cal A}(\bar D^0
\pi^0)~~~.
\eeq
For the $\bar D^* \rho$ amplitudes, which are characterized by three partial
waves with orbital angular momenta $\ell = 0$, 1, and 2, these relations hold
separately for each partial wave.  In what follows we shall assume a single
partial wave to dominate the process when discussing amplitude relations, but
will see in Sec.~V B that this is an oversimplification.

The magnitude of the $I = 3/2$ amplitude is obtained from the $B^+ \to \bar D^0
\pi^+$ partial width: $|A_{3/2}|^2 = \Gamma(\bar D^0 \pi^+)$.  The magnitude of
the $I = 1/2$ amplitude is obtained from the combination 
\beq
|A_{1/2}|^2 = \frac{3}{2} \left[ \Gamma(D^- \pi^+) + \Gamma(\bar D^0
\pi^0) \right] - \frac{1}{2} \Gamma(\bar D^0 \pi^+)~~~. 
\eeq
The relative phase $\delta_I$ between isospin amplitudes is given by
\beq
\cos \delta_I = \frac{3 \Gamma(D^- \pi^+) + \Gamma(\bar D^0 \pi^+) - 6
\Gamma(\bar D^0 \pi^0)} {4|A_{1/2} A_{3/2}|}~~~. 
\eeq
When only an upper bound on the color-suppressed rate is available, a useful
upper bound \cite{NS} on $\delta_I$ is
\beq 
\sin^2 \delta_I \le 9 \Gamma(\bar D^0 \pi^0)/ 2 \Gamma(\bar D^0 \pi^+)~~~.
\eeq
Similar expressions hold for $\bar D^* \pi$, $\bar D \rho$, and $\bar D^* \rho$
decays.

Decomposing the decay amplitudes in terms of tree ($T$), color-suppressed
($C$), and exchange ($E$) amplitudes \cite{GHLR}, one finds expressions in
correspondence with those in Sec.~III:
\beq
{\cal A}(\bar D^0 \pi^+) = T + C~~,~~~ {\cal A}(D^- \pi^+) = T + E~~,~~~ {\cal
A}(\bar D^0 \pi^0) = (C - E)/\s~~~.
\eeq
The amplitude $E$ is used here to describe either an exchange subprocess
$\bar b d \to \bar c u$, or rescattering from the tree-dominated process
$\bar b d \to \bar c u \bar d d$ through $d \bar d$ annihilation into a
flavor-SU(3) singlet state.  As in Sec.~III, we neglect electroweak penguins.

Here, as in the case of $D$ decays, the set $T$, $C$, and $E$ is overcomplete,
so we cannot extract independent information on the magnitude of $E$.  In
principle one could perform a calculation based on the factorization
assumption, as mentioned in Sec.~III, to relate the tree contribution $T$ in,
e.g., $B^0 \to D^- \pi^+$ to that in a semileptonic decay process such as $B^0
\to D^- \ell^+ \nu_\ell$.  One already knows that this calculation works
approximately \cite{BJ,JLRFM,Wein}.  It is also likely that $C$ and $T$ are
relatively real since neither involves the highly absorptive annihilation
process described by $E$. 

The value of $E$ is expected to be quite small for a couple of reasons.  First,
estimates based on either rescattering or interaction with the spectator quark
suggest that $E$ will be much smaller in the $B$ system than in the charm
system. Second, whereas $\tau(D^+)/\tau(D^0) \simeq 2.5$, indicating the
importance of spectator interactions or long-distance physics for charm, the
corresponding ratio for $B$ mesons is much closer to 1, since \cite{PDG}
$\tau(B^+) = (1.65 \pm 0.04) \times 10^{-12}$ s while $\tau(B^0) = (1.56 \pm
0.04) \times 10^{12}$ s.  This implies that we will have some difficulty
determining the phase of $E$ relative to that of $C$ and $T$.  One will have to
determine the relative contributions of $C$ and $T$ (through calculations such
as those suggested for charm decays in Sec.~III B), and the small $E$
contribution, if it is present, will then have to be extracted.  At present
this is not possible because none of the color-suppressed decays $B^0 \to (00)$
has been observed \cite{PDG,CLEOcs}. 

The relevant branching ratios and rates \cite{PDG} are summarized in Table VII.
In all cases the amplitude relations degenerate into a nearly straight line
since the square roots of the rates in Table VII all are consistent with
$[\Gamma(0+)]^{1/2} - [\Gamma(-+)]^{1/2} = [2 \Gamma(00)]^{1/2}$.  This point
is emphasized in Table VIII, where we also quote Suzuki's limits \cite{SuzDB}
on the relative phases of $I=1/2$ and $I = 3/2$ amplitudes \cite{cmt}. 
Comparison of the second and third columns of Table VIII allows one to see how
far above the lower isospin bound each $(00)$ color-suppressed mode lies.  The
decay $B^0 \to \bar D^0 \pi^0$ should appear at a level not much below its
present upper experimental bound. 

% This is Table VII
\begin{table}
\caption{$B^+$ and $B^0$ branching ratios and decay rates.}
\begin{center}
\begin{tabular}{c c c} \hline \hline
Mode & Branching ratio & Decay rate \\
     & (percent)       & ($\times 10^{8} {\rm s}^{-1}$) \\ \hline
\multicolumn{3}{c}{$B^+$ decays} \\ \hline
$\bar D^0 \pi^+$    & $0.53 \pm 0.05$ & $32 \pm 3$ \\
$\bar D^{*0} \pi^+$ & $0.46 \pm 0.04$ & $28 \pm 2.5$ \\
$\bar D^0 \rho^+$   & $1.34 \pm 0.18$ & $81 \pm 11$ \\
$\bar D^{*0} \rho^+$ & $1.55 \pm 0.31$ & $94 \pm 19$ \\ \hline
\multicolumn{3}{c}{$B^0 \to (-+)$ decays} \\ \hline
$D^- \pi^+$         & $0.30 \pm 0.04$ & $19.2 \pm 2.6$ \\
$D^{*-} \pi^+$      & $0.276 \pm 0.021$ & $17.7 \pm 1.4$  \\
$D^- \rho^+$        & $0.79 \pm 0.14$ & $51 \pm 9$  \\
$D^{*-} \rho^+$     & $0.67 \pm 0.33$ & $43 \pm 21$ \\ \hline
\multicolumn{3}{c}{$B^0 \to (00)$ decays} \\ \hline
$\bar D^0 \pi^0$    & $< 0.012$ & $< 0.77$ \\
$\bar D^{*0} \pi^0$ & $< 0.044$ & $< 2.8$ \\
$\bar D^0 \rho^0$   & $< 0.039$ & $< 2.5$ \\ 
$\bar D^{*0} \rho^0$ & $< 0.056$ & $< 3.6$ \\ \hline \hline
\end{tabular}
\end{center}
\end{table}

% This is Table VIII
\begin{table}
\caption{Comparison of amplitudes for $B \to \bar D^{(*)} + (\pi, \rho)$
decays.} 
\begin{center}
\begin{tabular}{c c c c c} \hline \hline
Mode            & $[\Gamma(0+)]^{1/2} - [\Gamma(-+)]^{1/2}$ &
$[2 \Gamma(00)]^{1/2}$ & $\delta_I {\rm ~(max)}$ & $[\Gamma(0+)]^{1/2}/
[\Gamma(-+)]^{1/2}$ \\
                & ($\times 10^{4} {\rm s}^{-1/2}$) &
$(\times 10^{4} {\rm s}^{-1/2})$ & (degrees) & \\ \hline
$\bar D \pi$    & $1.3 \pm 0.4$ & $< 1.2$ & 19 & $1.29 \pm 0.11$ \\
$\bar D^* \pi$  & $1.1 \pm 0.3$ & $< 2.4$ & 46 & $1.25 \pm 0.08$ \\
$\bar D \rho$   & $1.9 \pm 0.9$ & $< 2.2$ & 25 & $1.27 \pm 0.14$ \\
$\bar D^* \rho$ & $3.1 \pm 1.9$ & $< 2.7$ & 40 & $1.48 \pm 0.40$ \\
\hline \hline
\end{tabular}
\end{center}
\end{table}

The possibility of a small $E$ contribution with a highly absorptive phase
cannot be excluded.  Present data are consistent with $T$ and $C$ contributions
with no relative phase.  If $E$ is negligible and $C$ is real relative to $T$,
each process is characterized by $|C+T|/|T| = [\Gamma(0+)]^{1/2}/
[\Gamma(-+)]^{1/2}$.  The entries in the last column of Table VIII are
consistent with a universal value of $C/T = 0.27 \pm 0.06$. 
\bigskip

\leftline{\bf B.  Relative phases of $B \to VV$ amplitudes}
\bigskip

The CLEO \cn~\cite{CLEOVV} has presented evidence in the decays $B^0 \to D^{*-}
\rho^+$ and $B^+ \to \bar D^{*0} \rho^+$ for complex phases between helicity
amplitudes, and for the presence of more than one partial wave in these decays.
The helicity amplitudes $A_0$ and $A_{\pm 1}$ for $B \to VV$ decays are
expressible in terms of $\ell = 0$, 1, and 2 partial waves $S$, $P$, and $D$ as
\cite{DDLR} 
\beq
A_{\pm 1} = \sqrt{\frac{1}{3}}S \pm \sqrt{\frac{1}{2}}P + \sqrt{\frac{1}{6}}D
~~,~~~
A_0 = - \sqrt{\frac{1}{3}}S + \sqrt{\frac{2}{3}}D~~~.
\eeq
Here the amplitudes are normalized such that $|A_0|^2 + |A_1|^2 + |A_{-1}|^2 =
|S|^2 + |P|^2 + |D|^2 = 1$. While a full analysis of the CLEO results, which in
any case are preliminary, is beyond the scope of this note, we point out
several interesting features. 

1) The fraction of the decay which is longitudinal, $|A_0|^2$, is about 0.86 in
both $B^0 \to D^{*-} \rho^+$ and $B^+ \to \bar D^{*0} \rho^+$, indicating that
no individual partial wave dominates the decay. 

2) For $B^0 \to D^{*-} \rho^+$ the helicity amplitudes $A_1$ and $A_{-1}$ are
unequal (and have unequal phases), indicating the presence of a $P$-wave as
well as $S$ and/or $D$ wave components.  Only the amplitude $A_1$ has a
non-trivial phase with respect to $A_0$. 

3) For $B^+ \to \bar D^{*0} \rho^+$, the amplitudes $A_1$ and $A_{-1}$ are
consistent with being equal, indicating that no P-wave component may be needed. 
However, in this case both $A_1$ and $A_{-1}$ have non-trivial phases with
respect to $A_0$.

These results may indicate the presence of non-trivial final-state interactions
in the $B \to \bar D^* \rho$ decays, since such relative phases in helicity
amplitudes cannot arise at the level of weak amplitudes.  The dominant
subprocess in all these decays should be $\bar b \to \bar c u \bar d$, with a
vanishing weak phase. 
\bigskip

\centerline{\bf VI.  SUMMARY}
\bigskip

We have reviewed a number of cases, many of which were first pointed out by
Suzuki \cite{SuzVP,SuzPP,SuzDB}, in which non-trivial final-state interactions
manifest themselves in decay processes involving the release of up to a few GeV
of energy.  These include $J/\psi$ decays to pairs of light mesons, charmed
meson decays, and possibly $B \to \bar D^* \rho$.  We have argued that such
large final-state phases may also occur in penguin processes involving
$\bar b \to s$ transitions, especially those in which flavor-singlet mesons
like $\eta'$ are produced.  Such final-state phases may be useful in searching
for direct CP violation in decays like $B^0 \to K^+ \pi^-$ and $B^+ \to K^+
\pi^0$.  Our conclusion regarding the possibility of large final-state
phases is more optimistic than that of Ref.~\cite{SuzW}, where typical strong
phases of order $20^\circ$ have been estimated, as a result of our conjecture
that the $c \bar c$ annihilation process contributing to the penguin amplitude
can be highly absorptive.

A number of open experimental questions remain.  These can shed light on
whether there is a universal pattern giving rise to large final-state
interactions, or whether these effects must be studied on a case-by-case basis.

\begin{enumerate}

\item Although large final-state interactions have been demonstrated in
$J/\psi$ decays to $VP$ \cite{SuzVP} and $PP$ \cite{SuzPP} final states,
we do not yet know whether the same is true for $\psi'$ decays.  Observation
of the process $\psi' \to K^{*+} K^- + {\rm c.c.}$ and reduction in errors
on the branching ratios for $\psi' \to K^{*0} \bar K^0 + {\rm c.c.}$ and
$\psi' \to \omega \pi^0$ would help clarify this question.

\item We cannot yet choose between a resonant and non-resonant interpretation
of the large relative phase between $I = 1/2$ and $I = 3/2$ amplitudes in $D
\to \bar K \pi$ and $D \to \bar K^* \pi$ decays.  Although a $J^P = 0^+$
resonance decaying to $\bar K \pi$ has been seen near the $D$ mass
\cite{Aston}, there does not yet exist a candidate for a corresponding $0^-$
resonance decaying to $\bar K^* \pi$.  Moreover, such a resonance should not
couple appreciably to $\bar K \rho$ if the resonant interpretation is correct.
By comparing Figs.~3 and 4, we see that such a resonance should be more
prominent in the $\bar K^* \pi$ channel than the $0^+$ resonance is in the
$\bar K \pi$ channel, since its enhancement of the $I = 1/2$ amplitude relative
to the $I = 3/2$ amplitude is greater.  The absence of an appreciable
resonant or rescattering enhancement in the $\bar K \rho$ channel may provide
a clue to the space-time properties of the enhancement mechanism.

\item A number of tests of the factorization hypothesis in charmed particle
decays \cite{BSW,NS,Wein} are available.  With new large samples of charmed
particle decays becoming available from several sources (e.g., the CLEO
detector at Cornell and the FOCUS Collaboration at Fermilab), it might be worth
re-examining two-and three-body $D$ decays to see if these relations continue
to hold. 

\item Rare decays of $B$ mesons, as pointed out also elsewhere \cite{resc}, can
shed light on at least the magnitude, if not the phase, of final-state
interaction effects.  Such effects would be manifested, for example, as (a) an
observable CP-violating difference between the branching ratios for $B^+ \to
K^0 \pi^+$ and $B^- \to \bar K^0 \pi^-$, (b) an enhancement of the rate for
$B^+ \to K^+ \bar K^0$ and its charge-conjugate, and a possible CP-violating
asymmetry in these two rates, and (c) a branching ratio for $B^0 \to K^+ K^-$
above the level of a few parts in $10^8$.  We have pointed out that if large
final-state phases have a universal nature as a result of highly absorptive
processes, the CP asymmetries in $B^+ \to K^0 \pi^+$ and $B^+ \to K^+ \bar K^0$
may not be so large, and the rate enhancements in $B^+ \to K^+ \bar K^0$ and
$B^0 \to K^+ K^-$ may be a preferable means of displaying large final-state
interactions. 

\item The question of large final-state phases due to rescattering in $B$
decays to charmed particles remains open.  Such phases could arise as a
result of annihilation of light-quark pairs.  More information on such
processes will be forthcoming once the color-suppressed processes
$B^0 \to (\bar D^{(*)0} + (\pi^0 {\rm~or}~\rho^0)$ have been observed.
\end{enumerate}

\bigskip

\centerline{\bf ACKNOWLEDGEMENTS}
\bigskip

I would like to thank the Physics Department and the Theory Group at the
University of Hawaii for their hospitality during part of this work, and S.
Olsen, S. Pakvasa, S. F. Tuan, and H. Yamamoto for helpful conversations there.
Part of this investigation was performed during the Workshop on {\it
Perturbative and Non-Perturbative Aspects of the Standard Model} at St. John's
College, Santa Fe.  I would like to thank the organizer Rajan Gupta, as well as
the participants of the workshop, for providing a stimulating atmosphere. I am
grateful to Amol Dighe, Michael Gronau, and Matthias Neubert for collaborations
on topics closely related to the work described here, to Roy Schwitters for the
opportunity to present some of these results at a symposium at the University
of Texas, to J. Cumalat and J. Wiss for information related to Dalitz plots,
and to S. Stone for useful discussions. This work was supported in part by the
United States Department of Energy under Contract No. DE FG02 90ER40560. 
\bigskip
% \newpage

% Journal and other miscellaneous abbreviations for references
% Phys. Rev. D style
\def \ajp#1#2#3{Am.~J.~Phys.~{\bf#1}, #2 (#3)}
\def \apny#1#2#3{Ann.~Phys.~(N.Y.) {\bf#1}, #2 (#3)}
\def \app#1#2#3{Acta Phys.~Polonica {\bf#1}, #2 (#3)}
\def \arnps#1#2#3{Ann.~Rev.~Nucl.~Part.~Sci.~{\bf#1}, #2 (#3)}
\def \art{and references therein}
\def \b97{{\it Beauty '97}, Proceedings of the Fifth International
Workshop on $B$-Physics at Hadron Machines, Los Angeles, October 13--17,
1997, edited by P. Schlein}
\def \carg{{\it Masses of Fundamental Particles -- Carg\`ese 1996}, edited by
M. L\'evy \ite, NATO ASI Series B:  Physics Vol.~363 (Plenum, New York, 1997)}
\def \cmp#1#2#3{Commun.~Math.~Phys.~{\bf#1}, #2 (#3)}
\def \cmts#1#2#3{Comments on Nucl.~Part.~Phys.~{\bf#1}, #2 (#3)}
\def \corn93{{\it Lepton and Photon Interactions:  XVI International
Symposium, Ithaca, NY August 1993}, AIP Conference Proceedings No.~302,
ed.~by P. Drell and D. Rubin (AIP, New York, 1994)}
\def \cp89{{\it CP Violation,} edited by C. Jarlskog (World Scientific,
Singapore, 1989)}
\def \dpff{{\it The Fermilab Meeting -- DPF 92} (7th Meeting of the
American Physical Society Division of Particles and Fields), 10--14
November 1992, ed. by C. H. Albright \ite~(World Scientific, Singapore,
1993)}
\def \dpf94{DPF 94 Meeting, Albuquerque, NM, Aug.~2--6, 1994}
\def \efi{Enrico Fermi Institute Report No. EFI}
\def \el#1#2#3{Europhys.~Lett.~{\bf#1}, #2 (#3)}
\def \epjc#1#2#3{Eur.~Phys.~J.~C {\bf#1}, #2 (#3)}
\def \f79{{\it Proceedings of the 1979 International Symposium on Lepton
and Photon Interactions at High Energies,} Fermilab, August 23-29, 1979,
ed.~by T. B. W. Kirk and H. D. I. Abarbanel (Fermi National Accelerator
Laboratory, Batavia, IL, 1979}
\def \hb87{{\it Proceeding of the 1987 International Symposium on Lepton
and Photon Interactions at High Energies,} Hamburg, 1987, ed.~by W. Bartel
and R. R\"uckl (Nucl. Phys. B, Proc. Suppl., vol. 3) (North-Holland,
Amsterdam, 1988)}
\def \ib{{\it ibid.}~}
\def \ibj#1#2#3{{\it ibid.}~{\bf#1}, #2 (#3)}
\def \ichep72{{\it Proceedings of the XVI International Conference on High
Energy Physics}, Chicago and Batavia, Illinois, Sept. 6--13, 1972,
edited by J. D. Jackson, A. Roberts, and R. Donaldson (Fermilab, Batavia,
IL, 1972)}
\def \ijmpa#1#2#3{Int.~J.~Mod.~Phys.~A {\bf#1}, #2 (#3)}
\def \ite{{\it et al.}}
\def \jmp#1#2#3{J.~Math.~Phys.~{\bf#1}, #2 (#3)}
\def \jpg#1#2#3{J.~Phys.~G {\bf#1}, #2 (#3)}
\def \lkl87{{\it Selected Topics in Electroweak Interactions} (Proceedings
of the Second Lake Louise Institute on New Frontiers in Particle Physics,
15--21 February, 1987), edited by J. M. Cameron \ite~(World Scientific,
Singapore, 1987)}
\def \KEK#1{{\it Flavor Physics} (Proceedings of the Fourth International
Conference on Flavor Physics, KEK, Tsukuba, Japan, 29--31 October 1996),
edited by Y. Kuno and M. M. Nojiri, Nucl.~Phys.~B Proc.~Suppl.~{\bf 59},
#1 (1997)}
\def \ky85{{\it Proceedings of the International Symposium on Lepton and
Photon Interactions at High Energy,} Kyoto, Aug.~19-24, 1985, edited by M.
Konuma and K. Takahashi (Kyoto Univ., Kyoto, 1985)}
\def \mpla#1#2#3{Mod.~Phys.~Lett.~A {\bf#1}, #2 (#3)}
\def \nc#1#2#3{Nuovo Cim.~{\bf#1}, #2 (#3)}
\def \nima#1#2#3{Nucl.~Instr.~Meth.~A {\bf#1}, #2 (#3)}
\def \np#1#2#3{Nucl.~Phys.~{\bf#1}, #2 (#3)}
\def \npbps#1#2#3{Nucl.~Phys.~B (Proc.~Suppl.) {\bf#1}, #2 (#3)}
\def \pisma#1#2#3#4{Pis'ma Zh.~Eksp.~Teor.~Fiz.~{\bf#1}, #2 (#3) [JETP
Lett. {\bf#1}, #4 (#3)]}
\def \pl#1#2#3{Phys.~Lett.~{\bf#1}, #2 (#3)}
\def \plb#1#2#3{Phys.~Lett.~B {\bf#1}, #2 (#3)}
\def \pr#1#2#3{Phys.~Rev.~{\bf#1}, #2 (#3)}
\def \pra#1#2#3{Phys.~Rev.~A {\bf#1}, #2 (#3)}
\def \prd#1#2#3{Phys.~Rev.~D {\bf#1}, #2 (#3)}
\def \prl#1#2#3{Phys.~Rev.~Lett.~{\bf#1}, #2 (#3)}
\def \prp#1#2#3{Phys.~Rep.~{\bf#1}, #2 (#3)}
\def \ptp#1#2#3{Prog.~Theor.~Phys.~{\bf#1}, #2 (#3)}
\def \rmp#1#2#3{Rev.~Mod.~Phys.~{\bf#1}, #2 (#3)}
\def \rp#1{~~~~~\ldots\ldots{\rm rp~}{#1}~~~~~}
\def \si90{25th International Conference on High Energy Physics, Singapore,
Aug. 2-8, 1990}
\def \slc87{{\it Proceedings of the Salt Lake City Meeting} (Division of
Particles and Fields, American Physical Society, Salt Lake City, Utah,
1987), ed.~by C. DeTar and J. S. Ball (World Scientific, Singapore, 1987)}
\def \slac89{{\it Proceedings of the XIVth International Symposium on
Lepton and Photon Interactions,} Stanford, California, 1989, edited by M.
Riordan (World Scientific, Singapore, 1990)}
\def \smass82{{\it Proceedings of the 1982 DPF Summer Study on Elementary
Particle Physics and Future Facilities}, Snowmass, Colorado, edited by R.
Donaldson, R. Gustafson, and F. Paige (World Scientific, Singapore, 1982)}
\def \smass90{{\it Research Directions for the Decade} (Proceedings of the
1990 Summer Study on High Energy Physics, June 25 -- July 13, Snowmass,
Colorado), edited by E. L. Berger (World Scientific, Singapore, 1992)}
\def \stone{{\it B Decays}, edited by S. Stone (World Scientific,
Singapore, 1994)}
\def \tasi90{{\it Testing the Standard Model} (Proceedings of the 1990
Theoretical Advanced Study Institute in Elementary Particle Physics,
Boulder, Colorado, 3--27 June, 1990), edited by M. Cveti\v{c} and P.
Langacker (World Scientific, Singapore, 1991)}
\def \vanc{29th International Conference on High Energy Physics, Vancouver,
23--31 July 1998}
\def \yaf#1#2#3#4{Yad.~Fiz.~{\bf#1}, #2 (#3) [Sov.~J.~Nucl.~Phys.~{\bf #1},
#4 (#3)]}
\def \zhetf#1#2#3#4#5#6{Zh.~Eksp.~Teor.~Fiz.~{\bf #1}, #2 (#3) [Sov.~Phys.
-- JETP {\bf #4}, #5 (#6)]}
\def \zpc#1#2#3{Zeit.~Phys.~C {\bf#1}, #2 (#3)}


\begin{thebibliography}{99}
\bibitem{cmt} Using the upper bounds on the color-suppressed $(00)$ processes
and the $1 \sigma$ lower bounds on the $B^+$ decay rates in Table VII, we find
upper bounds for $\delta_I$ of $20^\circ$, $45^\circ$, $24^\circ$, and
$28^\circ$ for the four processes listed in Table VIII.  The corresponding
numbers quoted by the CLEO \cn~\cite{HNN} are $20^\circ$, $42^\circ$,
$25^\circ$, and $26^\circ$.
\end{thebibliography}
\end{document}